\documentclass{rrparticle}
\usepackage{graphicx}

\title{Dark solitons for an extended quintic nonlinear Schr{\"o}dinger equation:
Application to water waves at $kh=1.363$}

\author[1]{F. Tsitoura}
\author[2]{T. P. Horikis}
\author[1,*]{D. J. Frantzeskakis}

\affil[1]{Department of Physics, University of Athens, Panepistimiopolis,
Zografos, Athens 15784, Greece \\
$^*$Corresponding author's Email: {\em dfrantz@phys.uoa.gr}}
\affil[2]{Department of Mathematics, University of Ioannina, Ioannina 45110, Greece}


\begin{document}
\maketitle

\begin{abstract}

{\it Abstract.}  We study the existence, formation and dynamics of gray solitons for an extended quintic nonlinear
Schr{\"o}dinger (NLS) equation. The considered model finds applications to water waves, when the
characteristic parameter $kh$ -- where $k$ is the wavenumber and $h$ is the undistorted water's
depth -- takes the critical value $kh=1.363$. It is shown that this model admits approximate dark
soliton solutions emerging from an effective Korteweg-de Vries equation and that two types of
gray solitons exist: fast and slow, with the latter being almost stationary objects. Analytical
results are corroborated by direct numerical simulations.

\end{abstract}

\section{Introduction}

It is well known that the one-dimensional (1D) defocusing nonlinear Schr{\"o}di-\\nger (NLS)
equation governing the complex field $u=u(x,t)$, namely:
\begin{equation}
{\rm i} u_t + p_0 u_{xx} - q_0 |u|^2 u =0,
\label{cnls}
\end{equation}
with $p_0, q_0 \in \mathbb{R}_+$, supplemented with nonvanishing boundary conditions at infinity,
i.e., $|u|=u_0 \in \mathbb{R}$ for $|x|\rightarrow \infty$, possesses dark soliton solutions
\cite{zs}. These structures are localized density depressions (``notches'') off of a stable
continuous-wave (cw) background, associated with a phase jump at the density minimum, and are
called ``black'' or ``gray'' depending on whether the density minimum is zero or non-zero,
respectively. There exists a vast amount of theoretical work and experimental observations of dark
solitons in optical systems featuring a defocusing nonlinearity \cite{ref-Yuri} and Bose-Einstein
condensates (BECs) with repulsive interatomic interactions \cite{djf,a,siam} (see also the review~\cite{b}). 
Nevertheless, dark solitons 
also arise in a variety of other physical systems, such as discrete mechanical \cite{mech} and
electrical \cite{el} lattices, magnetic films \cite{magn}, complex plasmas \cite{plasma}, nematic
liquid crystals \cite{ass}, and others.

Dark solitons are also known to exist in other versions of the NLS equation, such as the one
featuring competing cubic and quintic nonlinearities \cite{bar,me,her,vvk}; the relevant, so-called
cubic-quintic NLS (cqNLS) model is of the form:
\begin{equation}
{\rm i} u_t + p u_{xx} - q |u|^2 u + r |u|^4 u =0,
\label{cqnls}
\end{equation}
with $p, q, r \in \mathbb{R}_+$. This model appears naturally in the context of nonlinear optics as
an approximation of a saturable defocusing nonlinearity, through its Taylor expansion up to second
order in the normalized intensity $I=|u|^2$ \cite{ref-Yuri}. Furthermore, Eq.~(\ref{cqnls}) also
finds applications in the context of atomic BECs, with the quintic term accounting for three-body
interactions, provided that losses due to such interactions may be neglected -- see, e.g.,
Refs.~\cite{tb1,tb2} and rigorous analysis in Ref. \cite{tb3}. Additionally, in the same context of
atomic BECs, the quintic term may appear generically in the case where weak deviations from
one-dimensionality are taken into regard \cite{gora} (see also Ref.~\cite{b1} for the case of
attractive BECs). It is also worth mentioning that a defocusing, purely quintic NLS model [i.e.,
$q=0$ and $r<0$ in Eq.~(\ref{cqnls})] has also been used as a mean-field model describing strongly
interacting 1D Bose gases and, particularly, the so-called Tonks-Girardeau gas of impenetrable
bosons \cite{k1}; for such systems, dark solitons of the defocusing quintic NLS were found in an
exact analytical form \cite{k2} and their dynamics in the trap was also investigated \cite{fkp}.

More recently, there has been an interest in dark solitons in the context of surface gravity water
waves: indeed, black \cite{chblack} and gray \cite{chgray} solitons were observed on the surface of
water in two seminal experiments performed in water wave tanks. Key to this achievement was the
conduction of the experiments in the regime where $kh<1.363$ (here, $k$ is the wavenumber and $h$
is the undistorted water depth): indeed, in this regime, the pertinent NLS equation governing
weakly nonlinear water waves \cite{mja}, becomes of the {\it defocusing} type. In this case, the cw
solution of the NLS model {\it is not} subject to the Benjamin-Feir instability (also known as 
``modulational instability'' (MI) \cite{bf}) and can, thus, support nonlinear excitations, such as
the black and gray solitons observed in the experiments. Notice that in the opposite case, where
$kh>1.363$, the NLS equation is of the {\it focusing} type, its cw solution is modulationally
unstable, and this gives rise (instead of dark solitons) to other localized structures observed in
experiments, such as bright solitons \cite{yuen} and rogue waves \cite{rw} (see also the recent review \cite{c}).

According to the above, it becomes clear that the value of $kh$ is of paramount importance, as it
controls both the nature of the underlying NLS model (focusing/defocusing) and the type of its
soliton solutions. At the critical value, $kh=1.363$, the coefficient of the cubic nonlinear term
of the NLS vanishes and, at this order of approximation, the model equation becomes a linear
Schr{\"o}dinger equation for a free particle, featuring only dispersion. Then, a natural question
is which types of localized water wave structures can be supported at $kh=1.363$. Obviously, to
address this problem, one should resort and analyze a model resulting at a higher-order of
approximation, thus incorporating higher-order corrections. Several such models have been derived
and studied in the literature \cite{rsj1,kaku,sed1,sl,agaf,grim}. All these models, are in fact
special cases of the following generic equation:
\begin{equation}
{\rm i} u_t + r_1 u_{xx} + r_2 |u|^2 u + r_3 |u|^4 u
+{\rm i} \left[ r_4 |u|^2 u_x + r_5 (|u|^2)_x u \right] =0,
\label{pa}
\end{equation}
with coefficients $r_i \in \mathbb{R}$ ($i \in \{1,2,\ldots,5\}$), which has the following
property. Equation~(\ref{pa}) is uniformly valid, in the sense that it reduces to the classical NLS
model in the limit $r_i \rightarrow 0$ for $i\ge 3$, and is valid in the case under consideration,
i.e., at the critical value $kh=1.363$, where the coefficient of the cubic nonlinear term, $|u|^2
u$, is $r_2=0$. It is important to note that, as shown in Ref.~\cite{parkes}, Eq.~(\ref{pa}) can
formally be derived from any dispersive nonlinear system via the derivative expansion method
\cite{rpm}. Furthermore, apart from the context of water waves, such an extended quintic NLS
equation [see Eq.~(\ref{pa}) for $r_2=0$] finds still another physical application, as a model for
wave propagation in a discrete nonlinear transmission line \cite{ken}.

Motivated by the above results, our aim is to study a variant of Eq.~(\ref{pa}) for $r_2=0$, namely
an extended quintic NLS model, and show that it supports robust gray solitons -- provided that the
model under consideration possesses a modulationally stable cw. In fact, we will analyze, as a
particular example, the model proposed by Johnson \cite{rsj1}, which -- according to the specific
values of the coefficients of the model -- indeed supports such a stable cw. Our methodology and
main results, as well as the organization of the presentation, are described as follows. In
Section~2, we present the model and reduce it to a form similar to that of Eq.~(\ref{pa});
differences of the relevant equation with that studied in Refs.~\cite{kaku,sed1,sl,grim}, are also
discussed. Then, the model under consideration is asymptotically reduced to a Korteweg-de Vries
(KdV) equation via the reductive perturbation method \cite{rpm}. The KdV soliton is then used to
find approximate gray soliton solutions. We show that there exist two types of such solitons, slow
and fast ones. The former feature a supersonic behavior -- and, thus, are not physically relevant.
The latter, which are particular to the case of the extended quintic NLS model under consideration,
features a rapid velocity thus suggesting that stationary objects may not occur in these
conditions. In Section~3, we present results of numerical simulations, which corroborate our
analytical predictions. Finally, in Section~4, we summarize our conclusions.

\section{The model and its analytical consideration}

\subsection{The extended quintic NLS model and its cw solution}

We start by presenting the model under consideration, derived by Johnson \cite{rsj1}:
\begin{equation}
{\rm i} u_t -\alpha_1 u_{xx} - \alpha_2 |u|^2 u + \alpha_3 |u|^4 u
+ {\rm i} \left[\alpha_4 |u|^2 u_x - \alpha_5 (|u|^2)_x u \right] - \alpha_6 u \psi_t,
\quad \psi_x=|u|^2.
\label{m1}
\end{equation}
Here, $u(x,t)$ is the complex amplitude of the gravity wave, and the coefficients $\alpha_i \in
\mathbb{R}_+$ ($i\in\{1,2,\ldots,6\}$). For $kh=1.363$, the coefficient of the cubic term $|u|^2 u$
becomes $\alpha_2=0$. In this case, introducing the transformations $x \rightarrow
\sqrt{\alpha_3/2\alpha_1}x$, $t \rightarrow \alpha_3 t$, and $\psi \rightarrow
\sqrt{2\alpha_1/\alpha_3} \psi$, we cast our model in the following form:
\begin{equation}
{\rm i} u_t -\frac{1}{2}u_{xx} + |u|^4 u
+ {\rm i} \left[\beta |u|^2 u_x - \gamma (|u|^2)_x u \right]- \delta u \psi_t =0,
\quad \psi_x=|u|^2,
\label{m2}
\end{equation}
where the coefficients in Eq.~(\ref{m2}) are given by:
\begin{equation}
\beta = \frac{\alpha_4}{\sqrt{2\alpha_1 \alpha_3}},
\quad
\gamma =\frac{\alpha_5}{\sqrt{2\alpha_1 \alpha_3}},
\quad
\delta = \alpha_6 \sqrt{\frac{2\alpha_1}{\alpha_3}},
\end{equation}
and are all positive real numbers. 

Next, employing the transformation:
\begin{equation}
u(x,t)=v(x,t)\exp[-{\rm i}\delta \psi(x,t)],
\label{tr}
\end{equation}
we can reduce Eq.~(\ref{m2}) to the more convenient local form:
\begin{equation}
{\rm i} v_t -\frac{1}{2}v_{xx} + \left(1+ \frac{\delta^2}{2} +\beta \delta  \right)|v|^4 v
+ {\rm i}\left[(\beta+\delta) |v|^2 v_x +
\left(\frac{\delta}{2} -\gamma \right) (|v|^2)_x v \right]=0,
\label{m3}
\end{equation}
which will be the model of our focus hereafter. Obviously, the original system is now 
uncoupled\footnote{The reduction of Eq.~(\ref{m1}) to Eq.~(\ref{m3}) was also discussed in Ref.~\cite{kaku},
but in a more involved fashion.} and $\psi_x=|v|^2$. Here it should be noted that the
model~(\ref{m3}), having the form of an extended quintic NLS, differs from other extended versions
of the NLS equation, relevant to water waves of finite depth (see, e.g., Refs.~\cite{dy,sed2}, as
well as discussion in Refs.~\cite{sed1,sl,grim}), which were used to predict bright and dark
solitons on the surface of water \cite{sed2}: indeed, the later models include also linear
higher-order terms (e.g., third-order dispersion), while Eqs.~(\ref{m1}) only incorporate
higher-order nonlinear terms.

At this point, it is important to mention the following. In Ref.~\cite{kaku}, as well as in the
later works \cite{sed1,sl}, an equation similar to Eq.~(\ref{m3}) was derived, namely:
\begin{equation}
{\rm i} u_t -q_1 u_{xx} - q_2 |u|^4 u
-{\rm i} \left[ q_3 |u|^2 u_x + q_4 (|u|^2)_x u \right] =0,
\label{kak}
\end{equation}
where coefficients $q_i \in \mathbb{R}_+$ ($i\in\{1,\ldots,4\}$). Here, one should notice the
difference in the signs of the coefficients -- and, most notably, the one of the quintic term --
resulting in controversial results regarding the stability of the cw solution: in the framework of
Eq.~(\ref{m1}) it is modulationally stable (see below), while in the case of Eq.~(\ref{kak}) it is
subject to modulational instability \cite{kaku,sed1,sl}. This latter conclusion is also in
agreement with Refs. \cite{agaf,grim}, which focused on the analysis of the canonical form of
Eq.~(\ref{kak}), corresponding to $q_4=0$. Nevertheless, as mentioned above, we will analyze
Eq.~(\ref{m3}) as a generic example of the more general model~(\ref{pa}), which can support robust
gray solitons.

Key to our analysis below is the fact that the cw solution of Eq.~(\ref{m3}), namely:
\begin{equation}
v=\rho_0 \exp(-i\omega_0 t),
\quad
\omega_0 = - \left(1+ \frac{\delta^2}{2} +\beta \delta  \right)\rho_0^4,
\label{cw}
\end{equation}
where $\rho_0$ is an arbitrary real constant, is modulationally stable (and thus gray solitons on
top of this cw can be supported). This can easily be confirmed via a standard MI analysis (see,
e.g., Ref.~\cite{ref-Yuri}), whereby the wavenumber $k$ and frequency $\omega$ of a small
perturbation $\propto \exp[{\rm i}(kx-\omega t)]$ satisfy the following dispersion relation:
\begin{equation}
\omega = - \frac{1}{2}k \left[\rho_0^2 (- 2\beta + 2\gamma - 3\delta)
\pm \sqrt{{k^2} + \alpha \rho_0^4}\right],
\end{equation}
where
\begin{equation}
\alpha = 4\gamma^2 +5 \delta^2 + 4(2+\beta \delta - 2 \gamma \delta).
\label{a}
\end{equation}

As long as $\alpha>0$, $\omega \in \mathbb{R}~\forall k \in \mathbb{R}$, which means that the cw is
modulationally stable. As we see below, the availability of coefficients $\alpha_i$
($i\in\{1,2,\ldots,6\}$) of Eq.~(\ref{m1}) \cite{rsj1}, which in turn provides the coefficients
$\beta$, $\gamma$, and $\delta$ in Eq.~(\ref{m3}), allows us to deduce that this is indeed the case:
 the parameter $\alpha$ is always positive, and thus the cw solution~(\ref{cw}) is modulationally stable
$\forall k\in\mathbb{R}$.

\subsection{Reduction to the KdV equation and soliton solutions}

Next, we apply the reductive perturbation method \cite{rpm} to derive from Eq.~(\ref{m3}) an
effective KdV equation; the soliton solution of the latter, will be then used to derive approximate
gray soliton solutions of Eq.~(\ref{m3}). Notice that a similar approach has been used in the past
in the context of nonlinear optics, where KdV-like gray solitons were obtained for NLS models
perturbed by higher-order effects --see, e.g., Refs.~\cite{d1,d2} where the analytical approach is
also detailed. We start by introducing in Eq.~(\ref{m3}) the Madelung transformation,
\begin{equation}
v(x,t) = \sqrt{\rho(x,t)}\exp[i\phi(x,t)],
\end{equation}
where the real functions $\rho$ and $\phi$ denote, respectively, the density and phase of the unknown
field $v$. 

Then, considering small-amplitude slowly-varying modulations of the cw~(\ref{cw}), we
look for solutions in the form of the following asymptotic expansions:
\begin{equation}
\rho= \rho_0 +\sum_{j=1}^{\infty}\epsilon^j \rho_j(X,T),
\quad
\phi= -\omega_0 t +\sum_{j=1}^{\infty}\epsilon^{j-1/2} \phi_j(X,T).
\label{exp}
\end{equation}
Here, $0<\epsilon \ll 1$ is a formal small parameter, and the unknown densities $\rho_j$ and phases
$\phi_j$ are functions of the slow variables:
\begin{equation}
X=\epsilon^{1/2}(x-ct), \quad T= \epsilon^{3/2} t,
\label{slowv}
\end{equation}
with $c$ being the so-called ``speed of sound'' (to be determined), namely the speed of the linear
waves propagating on top of the cw background. Notice that the specific form of the asymptotic
expansions~(\ref{exp}) as well as the choice of the slow variables~(\ref{slowv}) are such that
dispersive and nonlinear effects come into play at the same order of approximation (see
Ref.~\cite{rpm}).

At the lowest-order of the reductive perturbation technique, we arrive at a system of two linear
equations. The compatibility condition of these equations is the algebraic equation:
\begin{equation}
c^2 -(2\beta-2\gamma+3\delta)\rho_0 c +
(\delta^2 -2\gamma \delta -2 \beta \gamma + \beta \delta +\beta^2 -2)\rho_0^2=0,
\label{c2}
\end{equation}
from which we obtain the speed of sound:
\begin{equation}
c=c_\pm=\frac{1}{2}\rho_0 \left( 2\beta-2\gamma + 3\delta \pm \sqrt{\alpha} \right),
\label{c}
\end{equation}
where $\alpha$ is given in Eq.~(\ref{a}). We observe that the condition $\alpha>0$ ensuring that
$c\in\mathbb{R}$ is identical with the one concerning the stability of the cw solution.
Nevertheless, as mentioned above, $\alpha$ is positive, which dictates the existence of two speeds
of sound, a ``fast'' and a ``slow'' one, $c_+$ and $c_-$, respectively. In addition, at the same
order of approximation:
\begin{equation}
\phi_{1X}= \Delta \rho_1,
\quad
\Delta = -\frac{(2\beta \delta +\delta^2 +2)\rho_0}{c-(\beta+\delta)\rho_0 },
\label{con}
\end{equation}
which connects the unknown phase $\phi_1$ with the density $\rho_1$.

To the next order of approximation, we obtain two nonlinear equations. The compatibility conditions
of the later is again the algebraic equation~(\ref{c2}) and the KdV equation for the density
$\rho_1$:
\begin{equation}
A \rho_{1T} + B \rho_1 \rho_{1X} + \Gamma \rho_{1XXX} =0,
\label{kdv}
\end{equation}
where the coefficients $A$, $B$ and $\Gamma$ are given by:
\begin{align}
A &= \frac{4\rho_0[2c +(2\gamma -2\beta -3\delta)\rho_0]}{c-(\beta+\delta)\rho_0},
\\
B&= \frac{4\tilde{B}}{[c-(\beta+\delta)\rho_0]^2},
\\
\tilde{B}&=2c^3 -c^2\rho_0(3\beta -2\gamma+4\delta)+2c\rho_0^2(2+2\beta \delta +\delta^2)
\nonumber \\
&+\rho_0^3[\beta^3+2\gamma(2-\beta^2)-\delta (6+\delta^2)-\beta(4+3\delta^2)],
\\
\Gamma &= -\frac{c-(\beta -2\gamma +2\delta)\rho_0}{(2+2\beta \delta +\delta^2)\rho_0}.
\end{align}

The soliton solution of the KdV, Eq.~(\ref{kdv}), reads:
\begin{equation}
\rho_1 = \frac{12\kappa^2 \Gamma }{B} {\rm sech}^2\left[
\kappa \left(X-\frac{4 \kappa^2 \Gamma }{A} T\right)-X_0 \right],
\label{sol1}
\end{equation}
where $\kappa$ is an arbitrary real constant of order $O(1)$, and $X_0$ denotes the initial soliton
position. Using Eq.~(\ref{sol1}), we can also determine the phase through Eq.~(\ref{con}), namely
$\phi_1 =\Delta \int \rho_1(X',T){\rm d}X'$. To this end, using these expressions, we can write the
relevant approximate solution of Eq.~(\ref{m3}) in terms of the original variables as follows:
\begin{align}
v(x,t) &\approx  \sqrt{\rho_0 + \frac{12 \epsilon \kappa^2 \Gamma }{B} {\rm sech}^2(\xi)}
\exp\left[-{\rm i}\omega_0 t + \frac{12{\rm i} \epsilon^{1/2}\kappa \Gamma \Delta}{B}
\tanh(\xi) \right],
\label{sol2} \\
\xi&= \epsilon^{1/2} \kappa \left[x-c\left(1
+\frac{4\epsilon \kappa^2 \Gamma}{cA}\right)t-x_0 \right].
\label{xi}
\end{align}

As we will see below, here we have $B\Gamma<0$; this implies that the solution~(\ref{sol2}) is
characterized by a sech$^2$-shaped density dip, and a tanh-shaped phase jump across the density
minimum, thus having the form of a genuine gray soliton. At this point, it is important to point
out that, since coefficients $A$, $B$, $\Gamma$, and $\Delta$ are functions of $c$, which in turn
takes two distinct values, $c_\pm$ as per Eq.~(\ref{c}), the solution~(\ref{sol2}) describes
simultaneously two different types of gray solitons, namely fast and slow ones, corresponding to
$c_+$ and $c_-$, respectively.

\section{Numerical results}

Next, we compare our analytical findings with results of direct numerical simulations for
Eq.~(\ref{m3}). We start by considering the values of the parameters, as well as of the various
coefficients involved in our analytical approach. First of all, we use the values of parameters
$\alpha_j$ appearing in Eq.~(\ref{m1}) from Ref. \cite{rsj1}, corresponding to the case $kh=1.363$.
These values, in turn, lead to the following ones for the coefficients $\beta$, $\gamma$, and
$\delta$ appearing in Eqs.~(\ref{m2}) and (\ref{m3}):
\[
\beta=1.764,\; \gamma=1.147,\; \delta=2.329.
\]
Using the above, Eq.~(\ref{a}) dictates that parameter $\alpha$ takes the value: $\alpha=62.558
\rho_0^4>0$. Thus, the cw background is indeed not subject to MI, and undergoes a stable evolution.
This also suggests that there are two possible real solutions to Eq.~(\ref{c2}), namely:
\[
c_-=0.156\rho_0 \quad \text{and} \quad
c_+=8.065\rho_0,
\]
leading to two possible gray solitons, as mentioned above. It is clear that one of them is almost
stationary (the one corresponding to $c_-$), while the second moves substantially faster (the one
corresponding to $c_+$). This allows us to respectively term these two solutions ``slow'' and
``fast''.

\begin{figure}[h!tbp]
\centering
\includegraphics[width=0.65\linewidth]{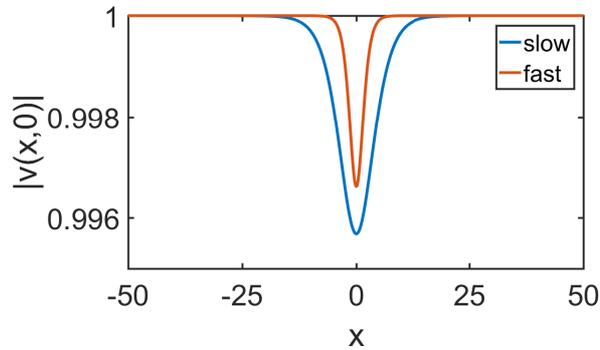}
\caption{
(Color online) The initial spatial profile of the fast (orange)
and slow (blue) gray solitons. Evidently, the former is shallower and
features a smaller width as compared to the latter. Parameter values are
$\rho_0=1$, while $\kappa=2$ for the slow solution and $\kappa=5$ for the fast one.
}
\label{fig1}
\end{figure}

Thanks to the nature of the KdV soliton (whereby amplitude, width and velocity are connected to
each other through the parameter $\kappa$), there exist also differences in the spatial profiles of
the slow and fast gray solitons. To better illustrate these differences, in Fig.~\ref{fig1} we plot
the two solutions at $t=0$. We choose $\rho_0=1$, $\epsilon=0.01$ and $\kappa=2$ for the slow
solution while $\kappa=5$ for the fast, so that the relative amplitudes (dips) are comparable. As
it is clearly seen, the slow solution is wider and deeper than the fast one, and it is thus
expected to propagate slower than the other, following the KdV dynamics above.

\begin{figure}[h!tbp]
\centering
\includegraphics[width=0.45\linewidth]{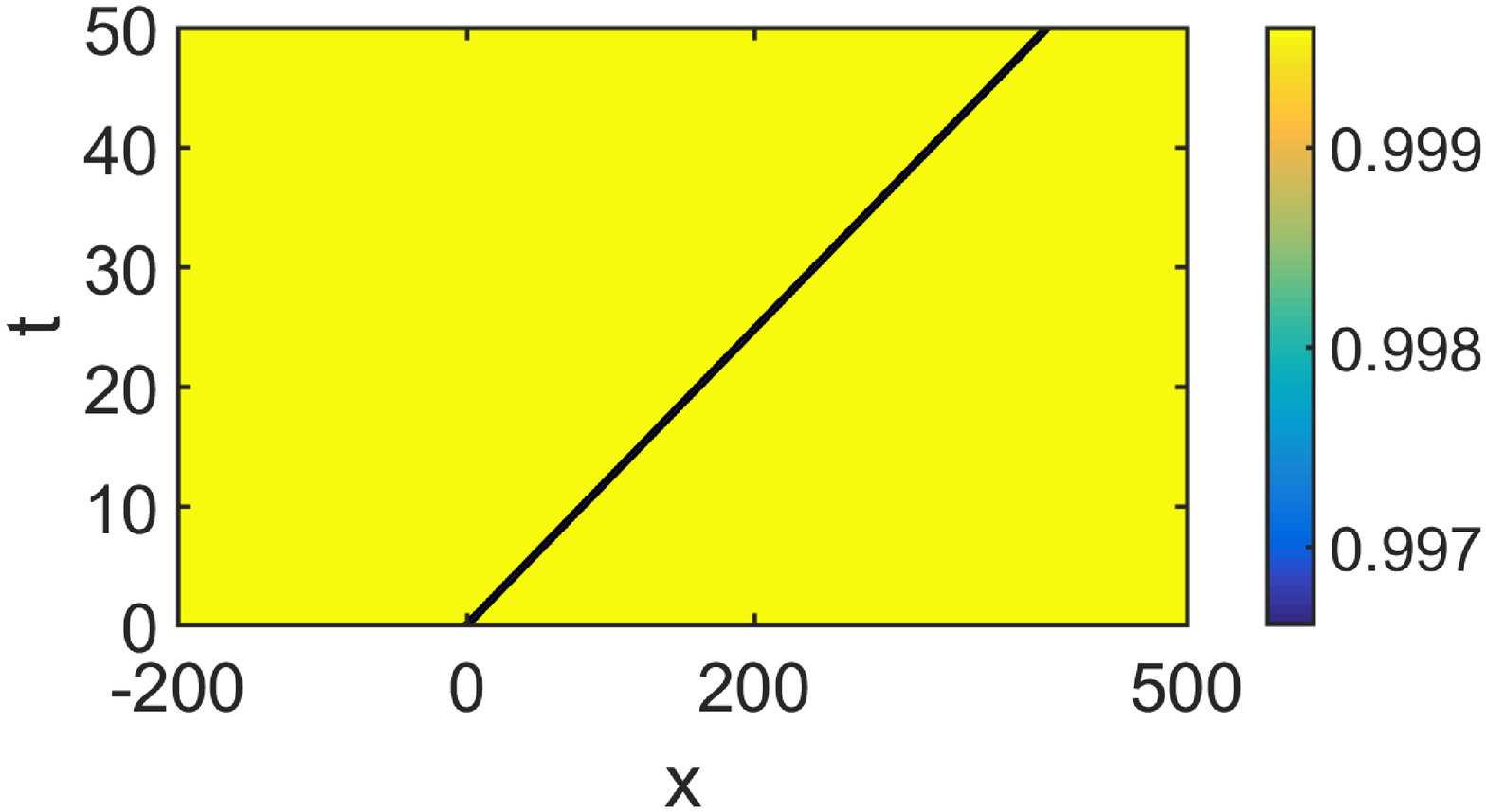}
\includegraphics[width=0.45\linewidth]{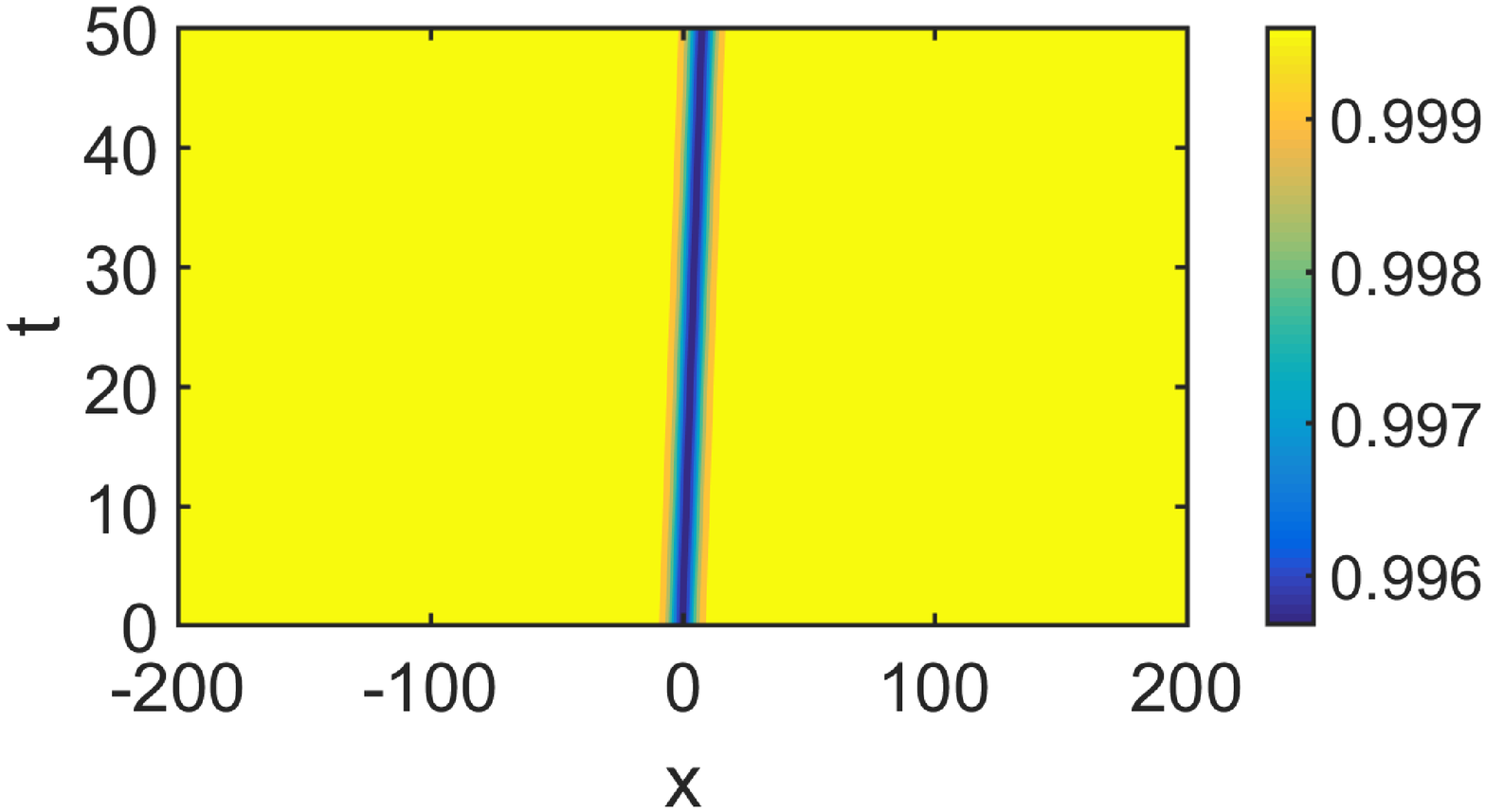} \\
\includegraphics[width=0.45\linewidth]{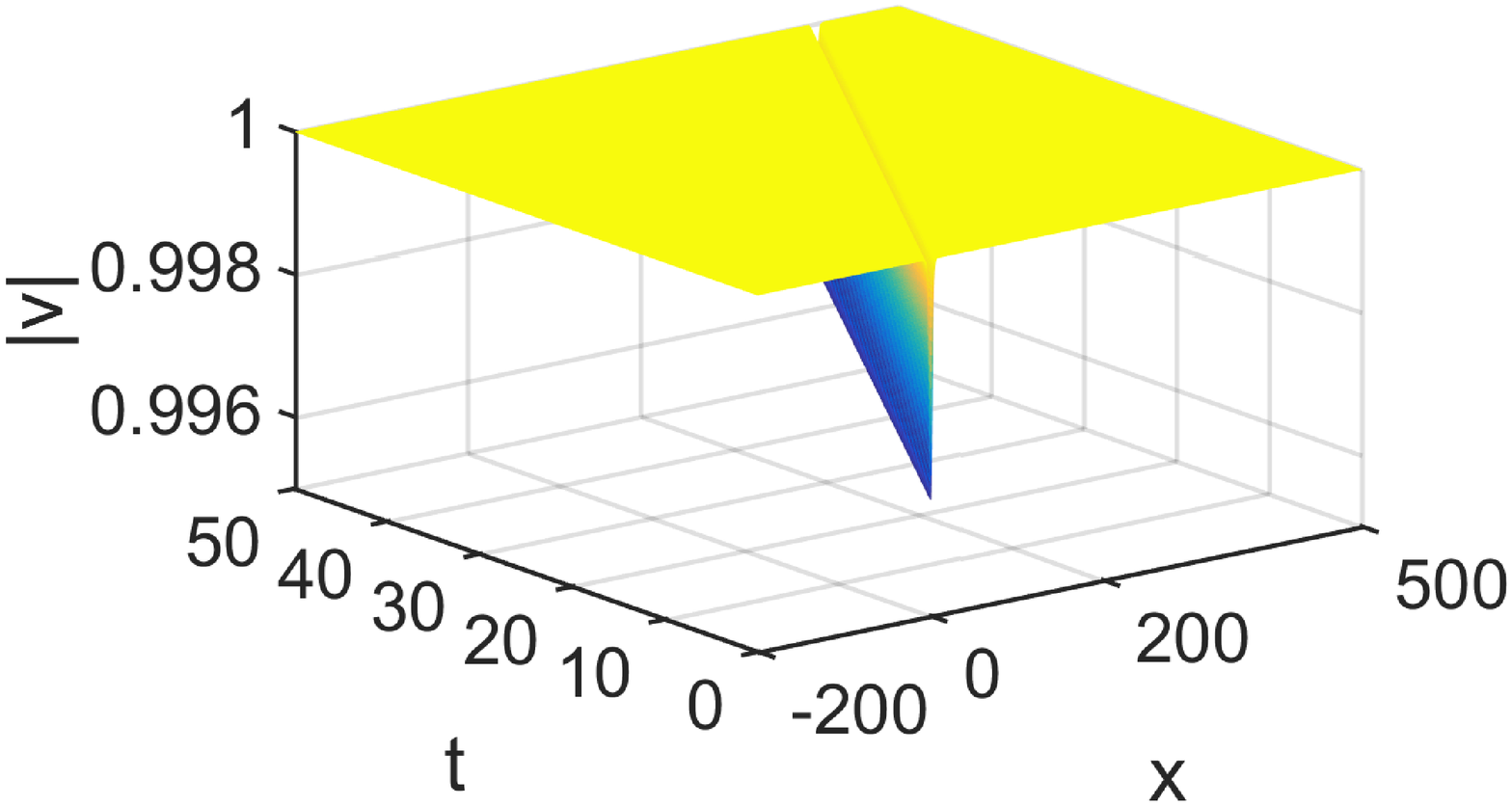}
\includegraphics[width=0.45\linewidth]{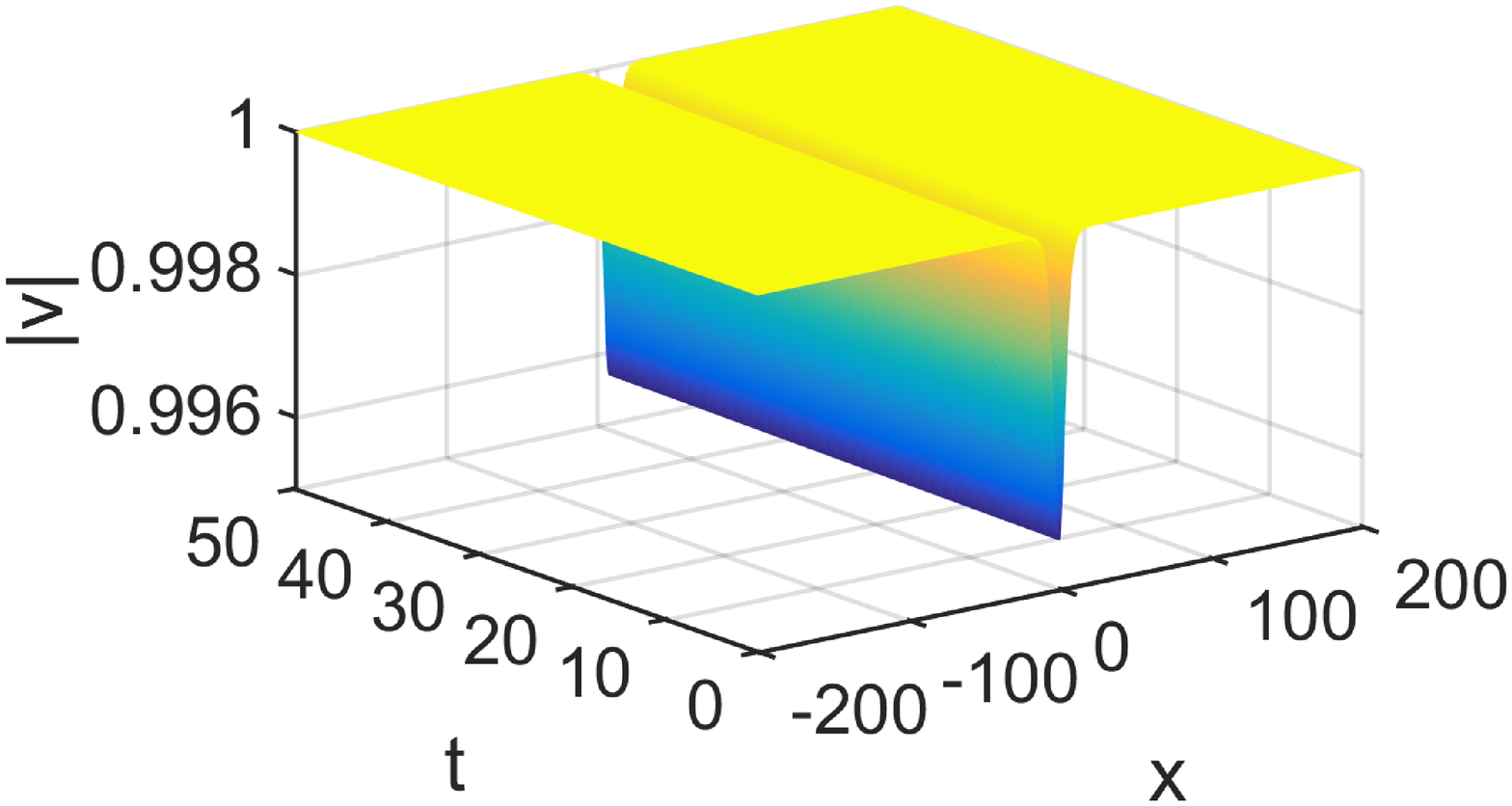} \\
\caption{
(Color online) Results of direct numerical simulations showing the evolution
of the fast (left column) and slow (right column) gray solitons
under the model Eq.~(\ref{m3}). Shown are contour plots (top panels)
and 3D plots (bottom panels) depicting the evolution of the modulus $|v(x,t)|$.
Parameter values are, again,
$\rho_0=1$ and $\kappa=2$ for the slow solution while $\kappa=5$ for the fast one.
}
\label{fig2}
\end{figure}

We have also performed direct numerical simulations. In particular, employing a fourth order
Runge-Kutta method, we have numerically integrated Eq.~(\ref{m3}) and used initial conditions (at
$t=0$) taken from Eqs.~(\ref{sol2})-(\ref{xi}). In Fig.~\ref{fig2}, we show the evolution of the fast
and slow solitons, corresponding to the initial conditions of Fig.~\ref{fig1}. The numerical results
agree with our analytical findings: indeed, both solutions follow the predicted KdV dynamics,
preserving their shapes, and traveling with constant velocity $C$. In the simulations, we find that
the latter is equal to:
$$C_{\rm fast}^{({\rm num})} = 8 \quad {\rm and} \quad C_{\rm slow}^{({\rm num})} = 0.15,$$
for the fast and slow solitons, respectively. On the other hand, according to the analytical
prediction, namely $C =c\left[1+4\epsilon \kappa^2 \Gamma/(cA)\right]$ (see Eq.~(\ref{xi})), the
respective velocities are found to be:
$$C_{\rm fast}^{({\rm an})} = 8.033 \quad {\rm and} \quad C_{\rm slow}^{({\rm an})} = 0.161.$$
Obviously, the agreement between numerical and analytical results is excellent. Nevertheless, it
should be noticed that our analysis dictates the existence of a supersonic gray soliton, with
$C_{\rm slow}^{({\rm an})}>c_-$, which can not occur in the realm of defocusing NLS models
\cite{ref-Yuri,djf}. The fact that slow solitons were not only found to exist, but also to be
robust in the simulations, implies that our analytical prediction for the solitons' velocities
slightly overestimates $C_{\rm slow}$.

In any case, thanks to the significantly small value of the velocity $c_-$ (and hence of $C_{\rm
slow}$), the slow gray soliton is almost stationary (at the reference system traveling with the
group velocity). This is a rather surprising result by itself for gray solitons, which are known to
be fast traveling objects \cite{ref-Yuri,djf}.

\section{Conclusions}

In conclusion, we have studied the existence, formation, and dynamics of gray solitons for an
extended quintic NLS equation. The considered model was the one proposed in Ref.~\cite{rsj1} for
surface water waves, when the parameter $kh$ takes the critical value $kh=1.363$. This model
possesses a stable cw solution, which contradicts later relevant results
\cite{kaku,sed1,sl,agaf,grim} dictating that the cw solution in this water wave problem (for
$kh=1.363$) is unstable. However, the considered model can be viewed as a paradigmatic example of a
general extended quintic NLS equation that can formally be obtained from any nonlinear dispersive
system \cite{parkes}; such a model, under certain conditions, can indeed support a stable cw
solution, on top of which gray solitons can occur.

Our analysis revealed that this is the case. Indeed, the model under consideration was studied
analytically by means of the reductive perturbation method, which led to an effective KdV equation.
The soliton solution of the latter, was then used to derive two types of approximate gray soliton
solutions of the original model: fast and slow ones. 

Direct numerical simulations have shown that
both types are particularly robust and evolve following closely the KdV dynamics. We have found
that the velocity of the slow soliton is quite small, such that this type of soliton can be deemed
to be almost stationary (in the reference frame moving with the group velocity), but with velocity
exceeding the speed of sound thus deeming these objects unphysical. This is a particular feature of
the extended quintic NLS equation under consideration, which, to the best of our knowledge, has no
analogue in other contexts where gray solitons also appear -- such as nonlinear optics
\cite{ref-Yuri} and Bose-Einstein condensates \cite{djf}.


\end{document}